\documentclass{PoS}
\usepackage{mathrsfs}
\title{Identification of Dark Matter with directional detection}

\ShortTitle{Identification of Dark Matter with directional detection}

\author{\speaker{J. Billard}, F. Mayet, D. Santos\\
        Laboratoire de Physique Subatomique et de Cosmologie, Universit\'e Joseph Fourier Grenoble 1,
  CNRS/IN2P3, Institut Polytechnique de Grenoble, Grenoble, France\\
        E-mail: \email{billard@lpsc.in2p3.fr}}

\abstract{ Directional detection is a promising search strategy to discover galactic Dark Matter.  
Taking advantage on the rotation of the Solar system around the Galactic center through the Dark Matter halo, it allows to show a direction 
dependence of WIMP events. Data of directional detectors are composed  of  energy and a 3D track 
for each  recoiling nuclei. Here, we present a Bayesian analysis method dedicated to data from upcoming directional detectors. 
However, we focus only on the angular part of the event distribution,
 arguing that the energy part of the background distribution is unknown. Two different cases are considered: a positive or a null detection of Dark Matter. In
 the first scenario, we will present a map-based likelihood method allowing to recover the main incoming direction of the signal and its significance, thus proving its
 Galactic origin. In the second scenario, a new statistical method is proposed. It is based on an extended likelihood in order to set robust and competitive exclusion limits.
  This method has been
 compared to two other methods and has been shown to be optimal in any detector configurations. Eventually, prospects for the MIMAC project are presented in the case of a
  10 kg $\rm CF_4$ detector with an exposition time of 3 years.}

\FullConference{Identification of Dark Matter 2010-IDM2010\\
		July 26-30, 2010\\
		Montpellier France}

\begin{document}
\section{Directional detection framework}
\begin{figure}[t]
\begin{center}
\includegraphics[scale=0.15,angle=90]{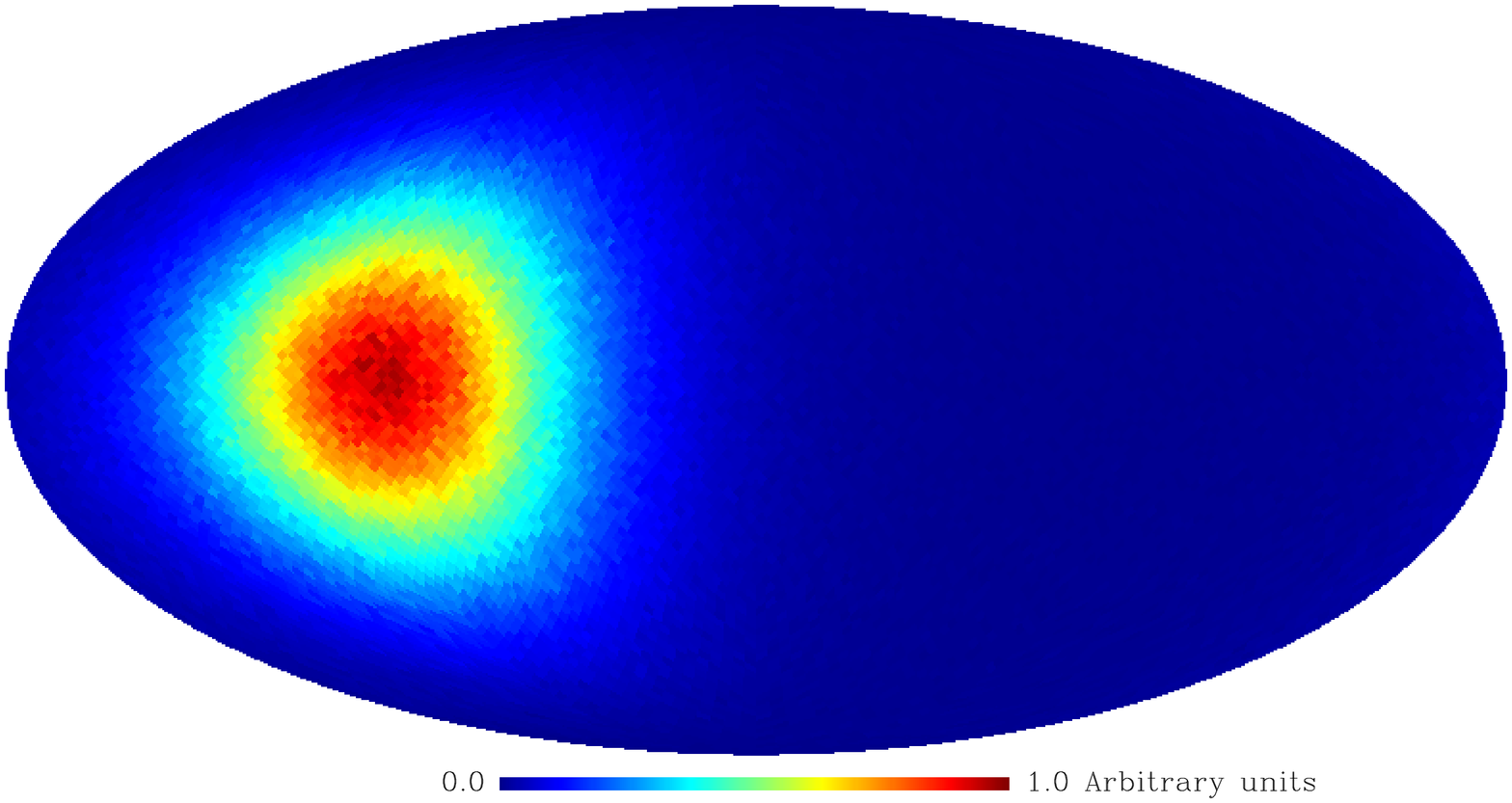}
\includegraphics[scale=0.15,angle=90]{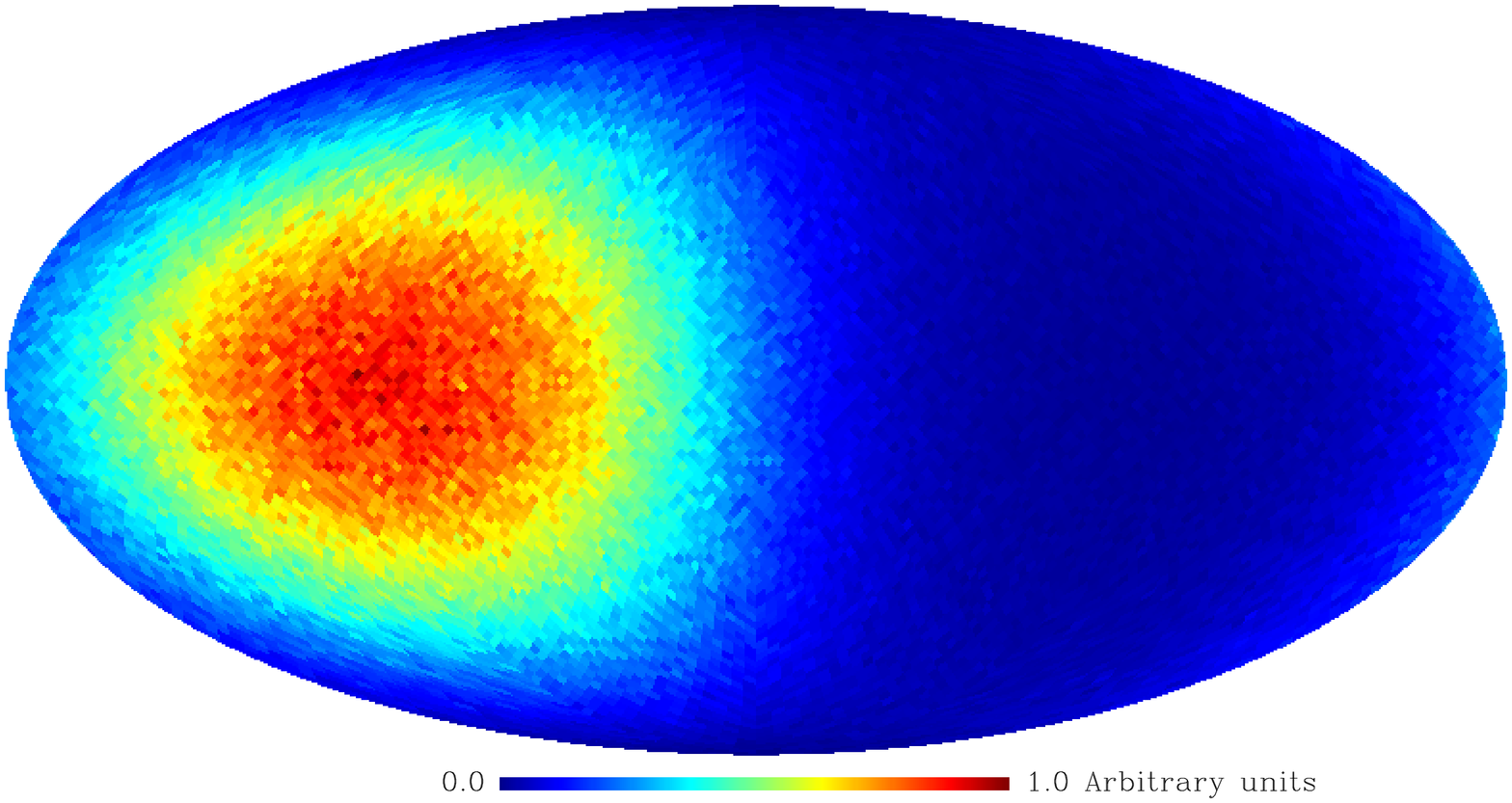}
\includegraphics[scale=0.15,angle=90]{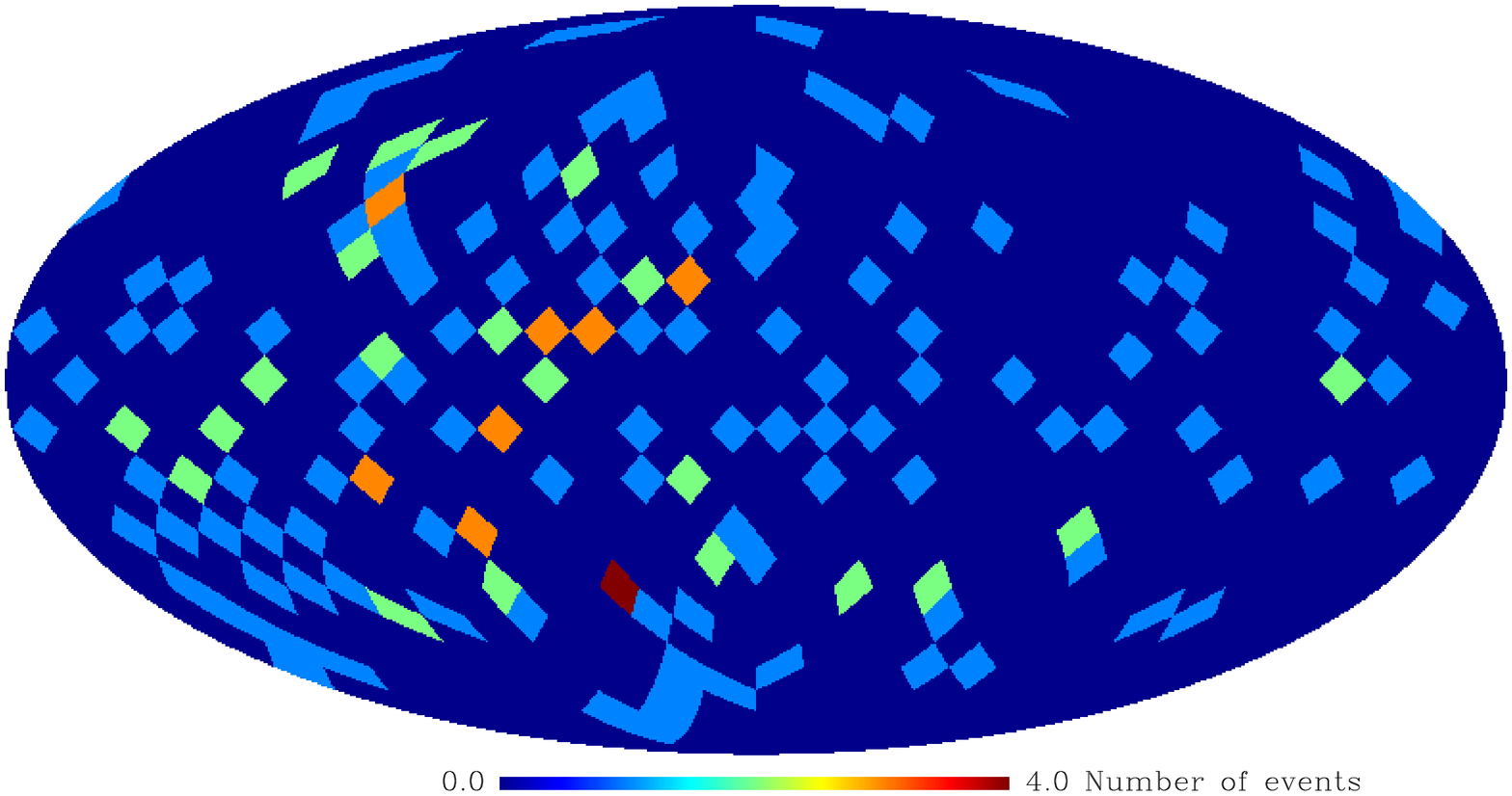}
\caption{From left to right : WIMP flux  in the case of an isothermal spherical halo,   WIMP-induced recoil distribution 
 and a typical simulated measurement :  100 WIMP-induced recoils and 100 background
events with a low angular resolution. Recoil maps are produced for a Fluorine target, a  100 GeV.c$^{-2}$ WIMP 
 and considering recoil energies in the range  5 keV $\leq E_R \leq$ 50 keV. Figures are taken from \cite{billard}.}  
\label{fig:DistribRecul}
\end{center}
\end{figure}

Taking advantage of the astrophysical framework, directional detection of Dark Matter is an interesting strategy in order to distinguish
 WIMP events from background ones.
Indeed, like most spiral galaxies, the Milky Way is supposed to be immersed in a halo of WIMPs which outweighs the luminous component by at 
least one order of magnitude. As the Solar System rotates around the galactic center through this Dark Matter halo, WIMPs should mainly come
 from the direction to which points the
Sun velocity vector and which happens to be roughly in the direction of the Cygnus constellation.
Then, a directional WIMP flux is expected to enter any terrestrial detectors (see fig.\ref{fig:DistribRecul} left) infering a directional
 WIMP
 induced recoil distribution which should be pointing toward the Cygnus Constellation, {\it i.e.} in the
  ($\ell_\odot = 90^\circ,  b_\odot =  0^\circ$) direction (see fig.\ref{fig:DistribRecul} middle).
  The latter corresponds to the expected WIMP signal probed by directional detectors and as it is shown on the fig.\ref{fig:DistribRecul}
  (middle), a strong anisotropy is expected \cite{spergel} while the background should be isotropic. Hence, we argue that a clear and unambigous identification of a Dark matter
  detection could be done by showing the correlation of the measured signal with the direction of the solar motion.\\
 
 Several project of directional detectors are being developed \cite{white,grignon} and some of them are already taking data \cite{DMTPC,DRIFT,NEWAGE}. We present a
 complete analysis framework dedicated to directional data. The first step when analysing directional data should be to looking for a signal pointing toward the
  Cygnus Constellation with a sufficiently high significance. If no evidence in favor of a Galactic origin of the signal is deduced from the previous
 analysis, then an exclusion limit should be derived. Here we will discussed the two different cases considering a 10 kg of CF$_4$ detector with a recoil energy range of 
 5 keV $\leq E_R \leq$ 50 keV.

\section{Case of positive detection}

An observed recoil map such as the one on the right of fig.\ref{fig:DistribRecul} could be obtained by a 10 kg CF$_4$ detector with
a WIMP-nucleon cross-section of $\sigma_n = 1.5\times10^{-3}$ pb and with a background rate of $\sim 0.07$ kg$^{-1}$.day$^{-1}$ in $\sim 5$ months exposition time.
 At first
sight, it seems difficult to conclude from this simulated recoil map that it does contain 
a fraction of WIMP events pointing towards the direction of the Solar motion. 
A likelihood analysis is developed in order to retrieve from a recoil map : the main direction of the incoming events in 
Galactic coordinates ($\ell, b$) and the number of WIMP events contained in the map. The likelihood value is estimated using a binned map 
of the overall sky with  Poissonian statistics,  as follows :
 \begin{equation}
 \mathscr{L}(m_\chi,\lambda, \ell,b) = \prod_{i=1}^{N_{\rm pixels}} P( [(1-\lambda) B_i + \lambda S_i(m_\chi ;\ell,b) ]|M_i)
 \end{equation}
where $B$ is the  background spatial distribution 
taken as isotropic, $S$ is the WIMP-induced recoil distribution and $M$ is the measurement. 
This is a four parameter likelihood analysis with $m_\chi$, 
 $\lambda = S/(B+S)$ the  WIMP fraction (related to the  background 
rejection power of the detector) and the coordinates ($\ell$, $b$) referring to the maximum of the 
WIMP event angular distribution.
Hence, $S(m_\chi;\ell,b)$ corresponds to a rotation of the $S(m_\chi)$ distribution 
by the angles ($\ell' = \ell - \ell_\odot$, $b' = b - b_\odot$). A scan of the four parameters with flat priors, allows to evaluate the likelihood between the measurement 
(fig.~\ref{fig:DistribRecul} right) and the theoretical distribution made of a superposition of 
an isotropic background and a pure WIMP signal (fig. \ref{fig:DistribRecul} middle). By scanning on $\ell$ and $b$ values, we ensure 
that there is no prior on the direction of the center of the WIMP-induced recoil distribution. As the observed map  is considered as a superposition of  
the background and the WIMP signal distributions, no assumption on the origin of each event is needed. Moreover, the likelihood method allows to recover $\lambda$,
 the WIMP fraction contained in the data. The advantage is twofold :
\begin{itemize} 
\item First, background-induced bias is avoided. This would not be the case with a method trying to 
evaluate a  likelihood on a map containing a fairly large number of background events considering only a pure WIMP reference distribution.  
\item Second, the value of $\lambda$ allows to access  the number of genuine WIMP events
and consequently to constrain the WIMP-nucleon cross-section as presented in \cite{billard} and to estimate the significance of the Dark Matter detection.
\end{itemize}
The four parameter likelihood analysis has been computed on 
the simulated map (fig. \ref{fig:DistribRecul} right) and the conclusion of this working example is the following; we found that 
it does contain a signal pointing towards the Cygnus constellation within 10$^\circ$, with $N_{\rm wimp}=106 \pm 17 \ (68 \% {\rm CL})$, corresponding to a high
significance detection of Galactic Dark Matter.\\

\begin{figure}[t]
\begin{center}
\includegraphics[scale=0.25,angle=270]{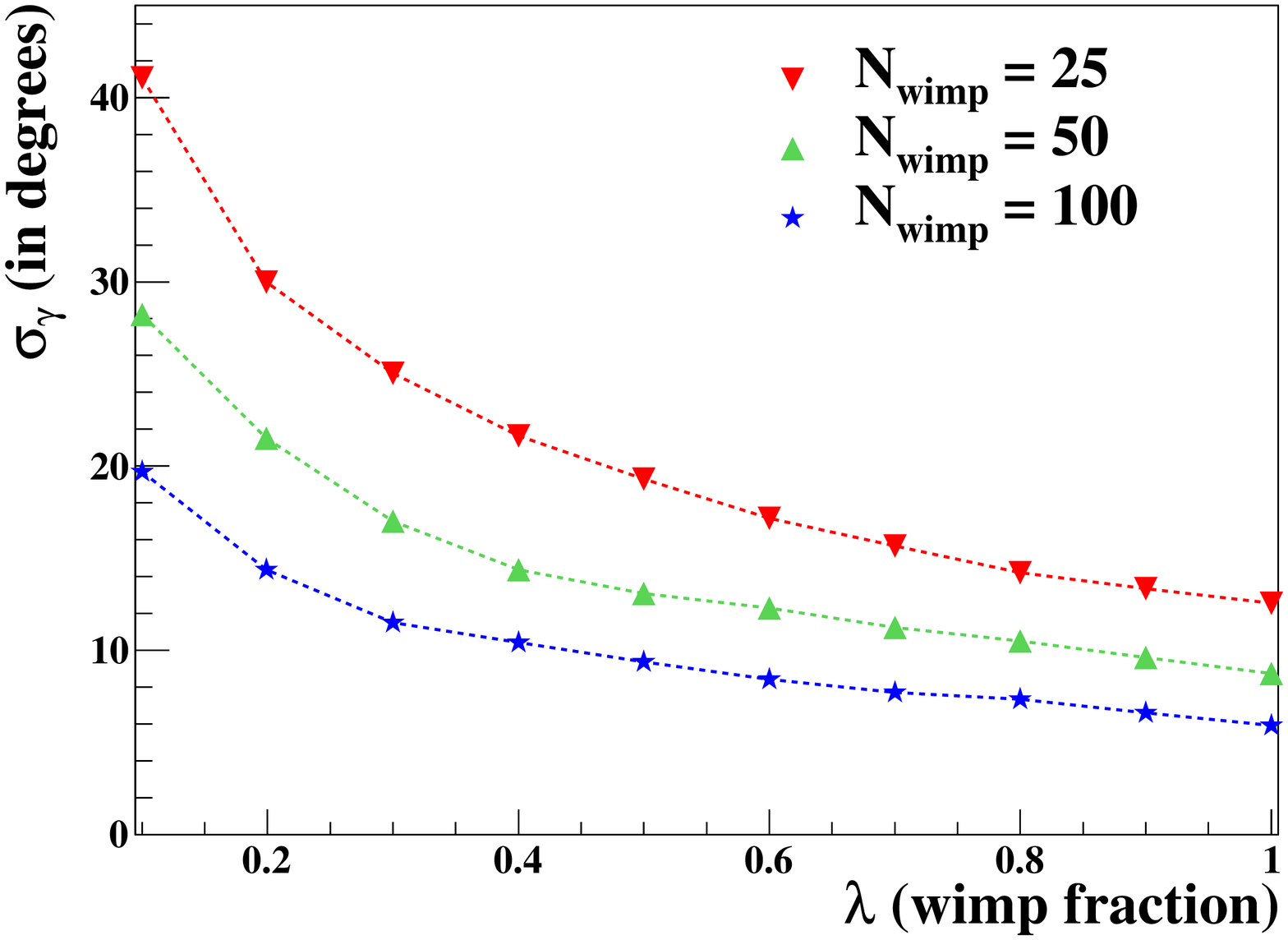}
\hspace{5mm}
 \includegraphics[scale=0.25,angle=270]{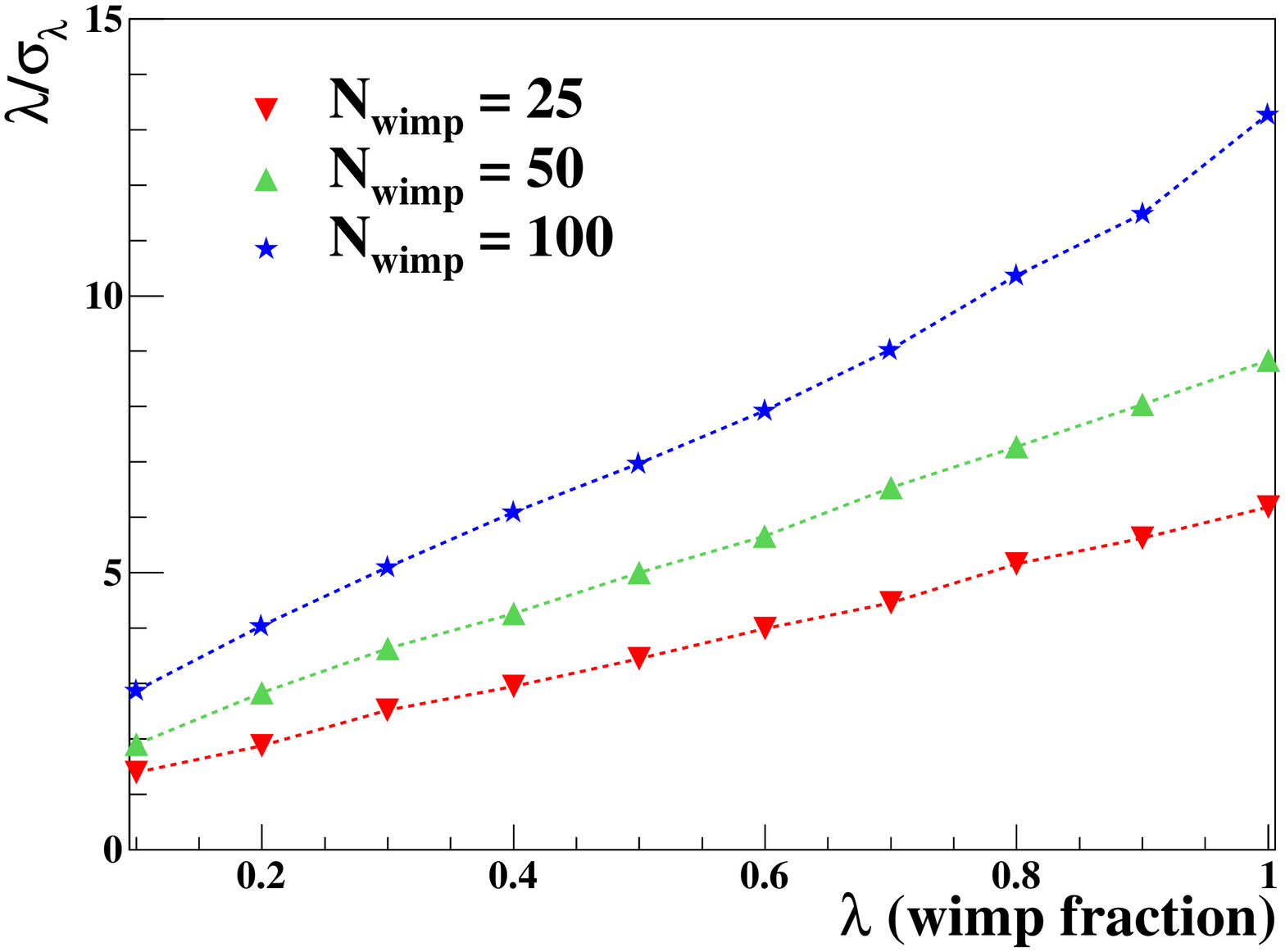}
\caption{Left panel presents the directional signature $\sigma_{\gamma}$ (in degrees) as a function of the WIMP fraction. 
Right panel presents the significance ($\lambda/\sigma_\lambda$) as a function of the WIMP fraction $\lambda=S/(S+B)$. 
Results are produced for a $^{19}$F target, a  100 GeV.c$^{-2}$ WIMP 
 and considering recoil energies in the range  5 keV $\leq E_R \leq$ 50 keV. Figures are taken from \cite{billard}.}  
\label{fig:exposition}
\end{center}
 \end{figure}

In order to explore the robustness of the method, and to ensure that it gives satisfactory results on a large range of exposure and background contamination, 
a systematic study has been
done with $10^4$ experiments for various number of WIMP events ($N_{\rm wimp}$) 
and several values of WIMP fraction in the observed map ($\rm \lambda$), ranging from 0.1 to 1. 
For a given cross-section, these two parameters are 
related respectively with the exposure and the rejection power of the offline analysis 
preceding  the likelihood method.\\
Figure~\ref{fig:exposition} presents on the left panel the directional signature, taken as the value of $\sigma_{\gamma} = \sqrt{  \sigma_\ell
\sigma_b }$,  the radius of the  $68 \%$ CL contour of the marginalised
$\mathscr{L}(\ell,b)$ distribution, as a function of $\lambda$. It is related to the ability to 
recover the main signal direction and to sign its Galactic origin. It can first be  noticed that the directional  
signature is of the order of 10$^\circ$ to 20$^\circ$ on a wide range of WIMP fractions.
Even for low number of WIMPs and for a low WIMP fraction (meaning a poor rejection power), 
the directional signature remains clear. From this, we conclude
that a directional evidence in favor of Galactic Dark Matter may be obtained with 
upcoming experiments even at low exposure (i.e. a low number of observed WIMPs) and with a non-negligible background 
contamination.\\
However, a convincing proof of the detection of WIMPs would require  a directional  
signature with  sufficient significance. We defined the significance of this identification strategy as $\lambda/\sigma_\lambda$, presented
on figure~\ref{fig:exposition} (right panel) as a function of $\lambda=S/(S+B)$. As expected, the significance is increasing both with the 
number of WIMP events and with the WIMP fraction, but we can notice that an evidence ($\rm 3 \sigma$) or a discovery ($\rm 5
\sigma$) of a Dark Matter signal would require either a larger number of WIMPs or a lower background contamination.\\
Using this map-based likelihood method, a directional detector may provide, as a first step, 
a Galactic signature  even with a low number of WIMPs. For instance, a signal pointing towards Cygnus within 
$20^\circ$ can be obtained with as low as 25 WIMPs with a 
50\% background contamination. For an axial cross-section on nucleon of  $\rm \sigma_{n} = 1.5 \times 10^{-3} \ pb$, this 
corresponds to an exposure of 400 kg.day in CF$_4$. In a second step, with an exposure four times larger, 
the directional signature would be only slightly better (10$^\circ$) but the significance would be 
much higher ($\sim 7\sigma$) and the detection much more convincing.

 \section{Case of a null detection}

This map-based likelihood analysis tool gives satisfactory results on a large range of exposure and background contamination levels. But, for very low number of
WIMP events and for very large background fractions ($\lambda \rightarrow 0$), when this method failed at recognizing a WIMP detection, obviously 
 an exclusion limit should be derived.  This will be the case for the very first results of directional
detectors with low exposures or if the WIMP-nucleon cross-section lies below $\rm \sim 10^{-5} \ pb$. Here, we present a Bayesian estimation of exclusion limits dedicated to directional data where only the angular part of the event distribution is
considered. The fact that both signal and background angular spectra are well known allows to derive upper limits using the Bayes' theorem. 
Considering an extended likelihood function with flat priors for both the expected number of WIMP events ($\mu_s$) and background events ($\mu_b$), and taking the evidence as  a normalization factor, it is reduced to 
$$
P(\mu_s,\mu_b|\vec{D}) \propto \frac{(\mu_s + \mu_b)^N}{N!}e^{-(\mu_s + \mu_b)}\times \prod_{i=1}^{N_{\rm pixels}} P\left( \frac{\mu_s}{\mu_s + \mu_b} S_i + \frac{\mu_b}{\mu_s + \mu_b}B_i|M_i\right)
$$
 where $\vec{D}$ refers to the characteristics of the data, as the total number of recorded events $N$, their direction and
their energy. It can be noticed that we use exactly the same likelihood definition than for the previous {\it discovery} method. Indeed, the $\lambda$ parameter has been
splitted into $\mu_s$ and $\mu_b$ and the WIMP mass is fixed. Hence, the probability density function of the parameter of interest $\mu_s$ can be
 derived by marginalizing $P(\mu_s,\mu_b|\vec{D})$ over the parameter $\mu_b$ and the excluded number of WIMP events $\mu_{\rm exc}$, corresponding to an excluded
 cross-section, at 90\% CL is obtained by solving:
\begin{equation}
\int_0^{\mu_{\rm exc}} P(\mu_s|\vec{D}) \ d\mu_s = 0.9,
\end{equation}
We have also used two other statistical methods: the Poisson method and the maximum gap method \cite{yellin1} applied to directional data. For each detector configurations and inputs, we have used 10 000 toy Monte
Carlo experiments in order to evaluate the frequency distributions of the excluded cross-section. Then, from each distribution, we can derive the median value of the 
excluded cross-section $\rm \sigma_{med}$ for the three statistical methods. More details may be found in \cite{billard2}.\\
We are interested on the effect of
detector configurations on the excluded cross-section. Indeed, even though several progresses have been done, 
sense recognition and angular resolution 
remain challenging experimental issues for directional detection of Dark Matter. \\

\begin{figure}[t]
\begin{center}
\includegraphics[scale=0.25,angle=270]{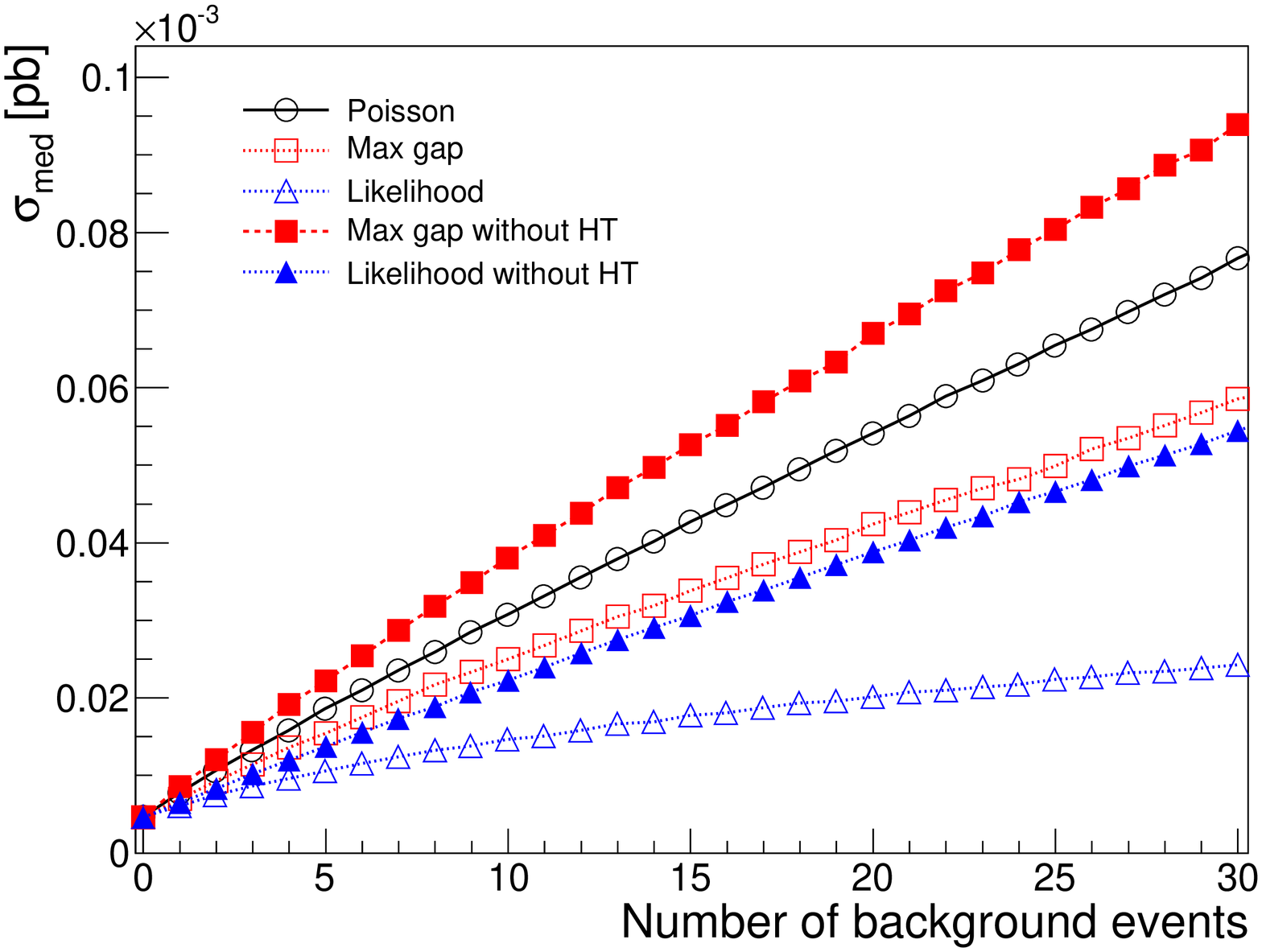}
\hspace{5mm}
\includegraphics[scale=0.25,angle=270]{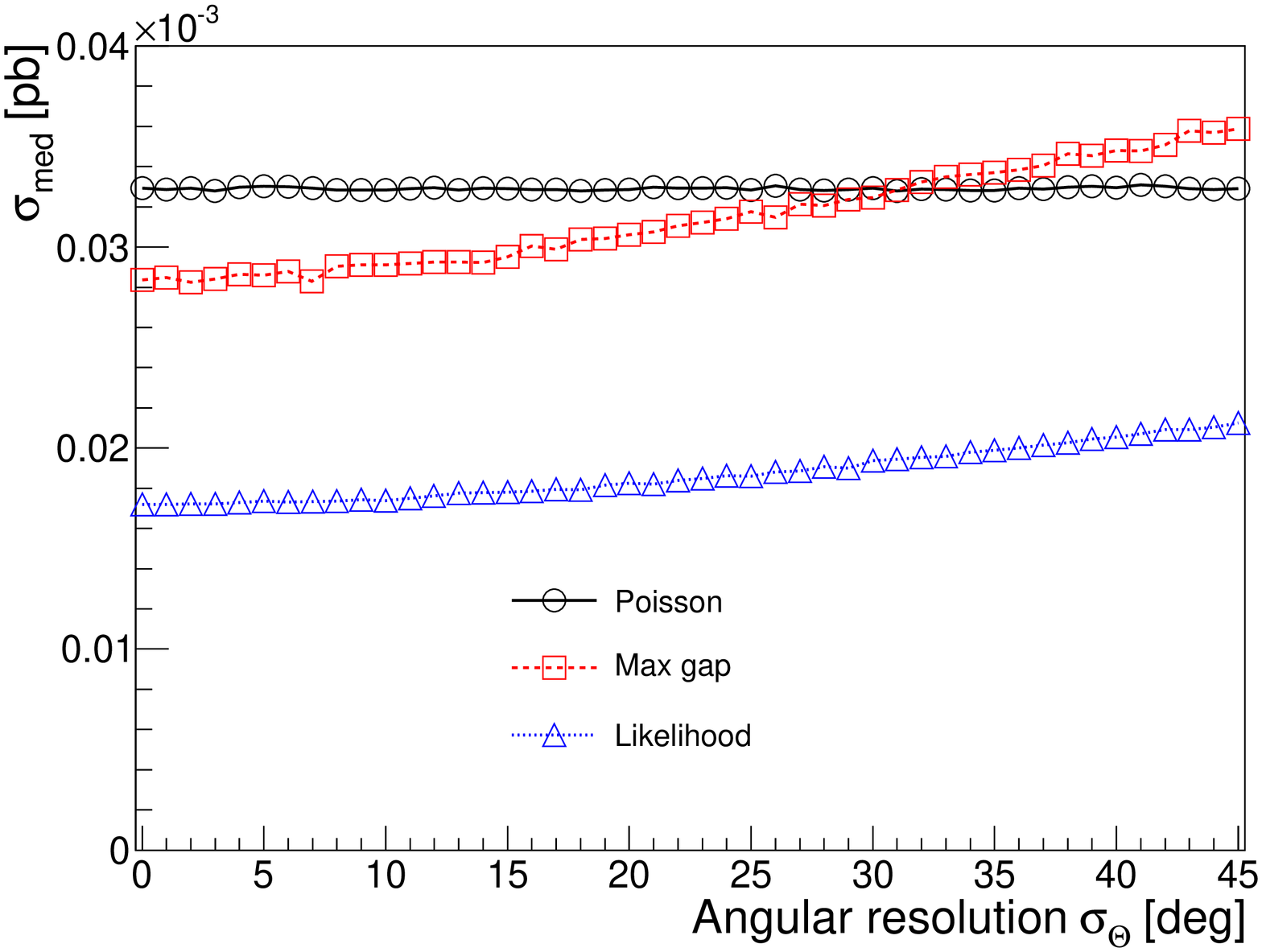}
\caption{The median upper limit cross section obtained by the three different statistical methods as a function of the number of background events in the case of pure
 background (on the left) and as a function of the angular resolution (on the right).
Left panel: The results without sense recognition are presented with filled markers while empty markers refer to a detector with sense recognition.
Right panel:  10 expected background events and one expected WIMP event. Figures are taken from \cite{billard2}.}  
\label{fig:BackgroundPureSanSHT}
\end{center}
\end{figure}

Without sense recognition, a recoil coming from ($\cos\gamma$,$\phi$) cannot be distinguished from a recoil coming
from ($-\cos\gamma$,$\phi + \pi$). The effect of no sense recognition in the case of pure background data is shown on figure \ref{fig:BackgroundPureSanSHT} where it is
compared with the case where the detector has a 100\% efficiency in the sense recognition. 
First of all, we can see on figure \ref{fig:BackgroundPureSanSHT} that the Likelihood method overcomes the two others in both cases. Secondly, 
 the absence of sense recognition only mildly alter the result: a factor of three at high background contamination.
 Taken at face value, this result suggests that sense recognition may not be so important for directional detection when setting exclusion limits. 
The difference between 100\% sense recognition on the whole recoil energy range, which is obviously unrealistic, and no sense recognition is only minor.
 The worst case is indeed a partial sense recognition strongly depending on the recoil energy. 
In this case, we suggest not to consider this information to set robust exclusion limits. However, sense recognition remains a key issue
worth investigating,  for WIMP discovery which is the ultimate goal of directional detection.\\

About the angular resolution, even if simulations
 show that straight line tracks may be 3D reconstructed with a rather small angular dispersion, realistic tracks in low
 pressure gaseous detectors would encounter a rather large angular dispersion. The lower is the recoil energy, the larger is the angular
 straggling. Hence, in the following 
we investigate the effect  on exclusion limits of using a detector with  a finite (realistic) angular resolution. Having a finite angular
resolution means that a recoil initially coming from the direction  $\hat{r}(\theta, \phi)$ is 
reconstructed as a recoil ${\hat{r}}^{\, \prime}(\theta^\prime, \phi^\prime)$ with a gaussian dispersion of 
 width $\sigma_\Theta$. Figure \ref{fig:BackgroundPureSanSHT} presents the cross section limit as a function 
of the angular resolution $\sigma_{\Theta}$ in the case of 10 expected background events and one expected WIMP
event. 
It can be noticed that the Maximum Gap and the likelihood methods are only slightly dependent on the angular 
resolution of the detector. The  deviation for the two methods is of the order of 30\ \%
from $\sigma_{\Theta} = 0^{\circ}$  to $\sigma_{\Theta} = 45^{\circ}$. Hence, as far as exclusion limits 
 are concerned, the effect of angular resolution is relatively small.\\ 
As a conclusion, this study outlines the need for detector commissioning, e.g. by using a neutron field. Indeed, in order to set coherent and robust upper
limits, it is necessary to have an accurate knowledge of the detector characteristics, even if the effect on exclusion limits is small.

\section{Prospects for the 10 kg CF$_4$ MIMAC detector}

\begin{figure}[t]
\begin{center}
\includegraphics[scale=0.4]{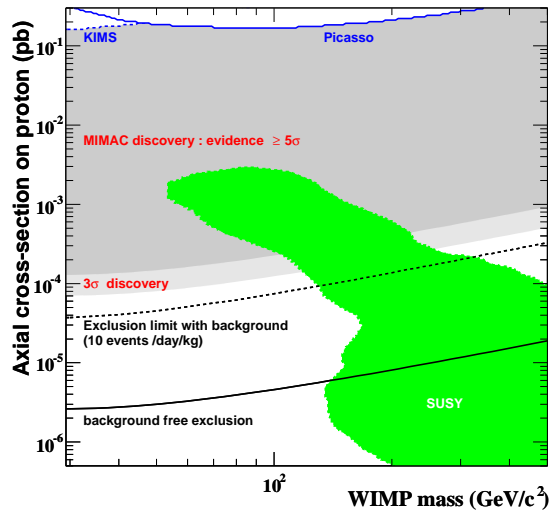}
\caption{Spin dependent cross-section on proton (pb) as a function of the WIMP mass ($\rm GeV/c^2$). 
Exclusion limits from some direct detection experiments are presented, KIMS~\cite{kims} and Picasso~\cite{picasso} as well as
the theoretical region, obtained within the framework of the constrained minimal supersymmetric model taken from \cite{superbayes}.
Contours corresponding to a significance greater than 3$\sigma$ and 5$\sigma$ are represented in grey and light grey.
The exclusion limit corresponding to pure background data is represented as the black dashed line and the detector sensitivity as the black solid line.}
\label{fig:discovery}
\end{center}
\end{figure}

To end-up this study, we present projected discovery regions and exclusion limits for a forthcoming directional detector proposed by the MIMAC collaboration \cite{grignon}. We consider 
a  10 kg $\rm CF_4$ detector 
operated during  $\sim 3$ years,  allowing 3D 
track reconstruction, with a $10^\circ$ angular resolution, a  recoil energy range 5-50 keV and with a realistic background rate of 10 evts/kg/year.
On figure \ref{fig:discovery} we have represented the two different possibilities for such a detector.
\begin{itemize}
\item The grey and light-grey shaded areas represent the contours where a
detection of Dark Matter would have a significance greater than 3$\sigma$ and 5$\sigma$. These contours are deduced from the map-based likelihood method and its estimation of
the significance. Then, as an illustration, if the WIMP-nucleon cross-section is about $10^{-4}$ pb with a WIMP mass of 100 GeV.c$^{-2}$, the detector would have a
 Dark Matter detection with a significance greater than 3$\sigma$. 
\item If the WIMP-nucleon cross-section is lower than $10^{-5}$ pb, then an exclusion limit is deduced using the extended likelihood function (black dashed line). As a benchmark
and to illustrate the effect of background on exclusion limits, we have also represented the detector sensitivity corresponding to the case where no events are recorded
(black solid line).
\end{itemize} 
Figure \ref{fig:discovery} presents exclusion limits from direct detection experiments, KIMS~\cite{kims} and Picasso~\cite{picasso} as well as
the theoretical region, obtained within the framework of the constrained minimal supersymmetric model taken from \cite{superbayes}. We can conclude that a directional
detector like MIMAC will cover an important region of interest worth being investigated.\\

This final study highlights the fact that directional detection should provide either an unambiguous signature in favor of a detection of Dark Matter if the cross section is
reachable by such a detector or to set robust and competitive exclusion limits.

\end{document}